\documentstyle[12pt]{article}
\newcommand{\su}{SUSY--SO(10)}
\newcommand{\tb}{\tan\beta}
\newcommand{\mb}{$m_b$}
\newcommand{\mt}{$m_\tau$}
\newcommand{\be}{\begin{equation}}
\newcommand{\ee}{\end{equation}}
\newcommand{\beq}{\begin{eqnarray}}
\newcommand{\eeq}{\end{eqnarray}}
\newcommand{\nn}{\nonumber}
\newcommand{\aoa}{\frac{\alpha_0}{\alpha}}
\newcommand{\ovM}{{\overline M}}
\newcommand{\ovm}{{\overline m}}
\newcommand{\cO}{{\cal O}}

\def\gtap{\ \raisebox{-.4ex}{\rlap{$\sim$}} \raisebox{.4ex}{$>$}\ }

\def\npb#1#2#3{    {\it Nucl. Phys. }{\bf B#1} (19#2) #3}
\def\plb#1#2#3{    {\it Phys. Lett. }{\bf B#1} (19#2) #3}
\def\prd#1#2#3{    {\it Phys. Rev. }{\bf D#1} (19#2) #3}
\def\prl#1#2#3{    {\it Phys. Rev. Lett. }{\bf #1} (19#2) #3}
\def\zpc#1#2#3{    {\it Z. Phys. }{\bf C#1} (19#2) #3}
\begin{document}
\vspace*{10pt}
\hspace*{14.0em}\[
\hspace*{14.0em}\begin{array}{l}
\hspace*{14.0em}\mbox{MPI--PhT/95--100}\\
\hspace*{14.0em}\mbox{DFTT 63/95}\\
\hspace*{14.0em}\mbox{LBL-37884}\\
\hspace*{14.0em}\mbox{UCB-PTH-95/34}\\
\hspace*{14.0em}\mbox{October 1995}
\hspace*{14.0em}\end{array}\]
\vspace*{10pt}
\thispagestyle{empty}
\begin{center}
{\sc VIABLE $t$-$b$-$\tau$ YUKAWA UNIFICATION}\\
{\sc IN SUSY SO(10)}\footnote{This work was supported in part by the
Director, Office of
Energy Research, Office of High Energy and Nuclear Physics, Division of
High Energy Physics of the U.S. Department of Energy under Contract
DE-AC03-76SF00098, by the National Science Foundation under
grant PHY-90-21139, by the Polish Committee for Scientific Research
and by the EU grant ``Flavourdynamics''.}
\end{center}
\vspace{20pt}
\begin{center}
Hitoshi Murayama$^{1,2)}$,
Marek Olechowski$^{3)}$\footnote{
On leave of absence from the Institute of Theoretical Physics,
Warsaw University, Poland.}
and Stefan Pokorski$^{4)\dagger}$
\end{center}
\vspace{20pt}
\begin{center}
\it
$^1$ Department of Physics, University of California\\
Berkeley, CA 94720, USA\\
$^2$ Theoretical Physics Group, Lawerence Berkeley Laboratory\\
University of California, Berkeley, CA 94720, USA\\
$^3$ INFN, Sezione di Torino, Via P. Giuria 1, 10125 Turin, Italy\\
$^4$ Max-Planck Institute for Physics\\
F\"ohringer Ring 6, 80805 Munich, Germany
\end{center}
\vspace{45pt}
Abstract:
The supersymmetric SO(10) GUT with $t$--$b$--$\tau$ Yukawa coupling
unification has problems with correct electroweak symmetry breaking,
experimental constraints (especially $b\rightarrow s\gamma$)
and neutralino abundance, if the scalar masses
are universal at the GUT scale.  We point out that
non-universality of the scalar masses at the GUT scale generated both by
(1) renormalization group running from the Planck scale to the GUT scale
and (2) $D$--term contribution induced by the reduction of the rank of
the gauge group, has a desirable pattern to make the model
phemenologically viable (in fact the only one which is consistent with
experimental and cosmological constraints).
At the same time the top quark mass has to be either close to its quasi
IR--fixed point value or below $\sim$170 GeV. We also briefly discuss
the spectrum of superpartners which is then obtained.
\newpage
\renewcommand{\thepage}{\roman{page}}
\setcounter{page}{2}
\mbox{ }

\vskip 1in

\begin{center}
{\bf Disclaimer}
\end{center}

\vskip .2in

\begin{scriptsize}
\begin{quotation}
This document was prepared as an account of work sponsored by the United
States Government. While this document is believed to contain correct
information, neither the United States Government nor any agency
thereof, nor The Regents of the University of California, nor any of their
employees, makes any warranty, express or implied, or assumes any legal
liability or responsibility for the accuracy, completeness, or usefulness
of any information, apparatus, product, or process disclosed, or represents
that its use would not infringe privately owned rights.  Reference herein
to any specific commercial products process, or service by its trade name,
trademark, manufacturer, or otherwise, does not necessarily constitute or
imply its endorsement, recommendation, or favoring by the United States
Government or any agency thereof, or The Regents of the University of
California.  The views and opinions of authors expressed herein do not
necessarily state or reflect those of the United States Government or any
agency thereof or The Regents of the University of California and shall
not be used for advertising or product endorsement purposes.
\end{quotation}
\end{scriptsize}

\vskip 2in

\begin{center}
\begin{small}
{\it Lawrence Berkeley Laboratory is an equal opportunity employer.}
\end{small}
\end{center}

\newpage
\renewcommand{\thepage}{\arabic{page}}
\setcounter{page}{1}
Grand Unification (GUT) has been regarded as a serious candidate of
physics beyond the weak-scale to explain various puzzling features of
the standard model.  While the three gauge coupling constants fail
to unify in its simplest version, they meet at a scale
$M_{GUT} \simeq 2\times 10^{16}$~GeV in its supersymmetric (SUSY)
extension. SUSY-SO(10) GUT offers further exciting possibility that all
three Yukawa coupling constants of top, bottom quarks and tau lepton may
also unify at the same scale where the gauge coupling constants unify.
This is possible because the supersymmetric standard model contains two
Higgs doublets, and large values of $\tan\beta = v_2/v_1$ (the ratio of
the vacuum expectation values for the doublet $H_2$ and $H_1$ which
couple to the up and down quarks, respectively) lead to a proper bottom
top quark mass hierachy, with approximately equal b and t Yukawa
couplings \cite{BANKS}.
The consequence of such an exact unification of couplings is that the
top quark mass, $m_t$, and $\tb$ are determined, once the bottom quark
mass, $m_b$, the tau lepton mass, \mt, and the strong gauge coupling,
$\alpha_3$, are fixed
\cite{WHO}, \cite{CAOLPOWA}, \cite{HARASA}.

In this context, an interesting question is the issue of the
compatilibity of this exact Yukawa coupling unification with the
possibility of breaking the electroweak gauge symmetry through
radiative effects.  This question has been investigated in a
number of papers in the minimal \su~ models with universal
\cite{UNIV}, \cite{BOTUP}, \cite{CAOLPOWA}
and non-universal
\cite{NONUNIV1}--\cite{NONUNIV2}
soft supersymmetry breaking
parameters at the GUT scale. Moreover, it has been recently
observed that for these large values of $\tb$, potentially large
corrections to \mb~ may be induced through the supersymmetry
breaking sector of the theory
\cite{HARASA}, \cite{CAOLPOWA}.
Altogether, the requirement of a physically acceptable value for
the \mb~ and of the consistency with the recent CLEO result for
the decay $b\rightarrow s\gamma$ and with the condition $\Omega h^2<1$
for the relic abundance of the LSP strongly constrain the
minimal \su~ with radiative electroweak symmetry breaking. In
recent papers
\cite{NONUNIV2}, \cite{BOP}
it has been shown that those constraints rule
out the model with universal soft supersymmetry breaking terms
at the GUT scale and select certain class of non--universal
boundary conditions which lead to radiative
breaking with $M_2\gg\mu$, i.e.\ with higgsino--like lightest
neutralino, as the only acceptable scenario for the minimal SUSY-SO(10).

 From the theoretical point of view, non--universal SUSY breaking
terms at the GUT scale appear at present as a realistic
possibility. In  GUT models, even with universal boundary conditions
at the Planck scale, the renormalization group (RG) running to the GUT
scale generically leads to some non-universality of the scalar masses
at that scale
\cite{PP}, \cite{GATO}.
In addition, in models like SO(10), the reduction of the
rank of the gauge group by one at $M_{GUT}$, together with non-universal
scalar masses, generates additional non-universal contributions given
by the $D$-term of the broken U(1)
\cite{DREES}, \cite{HIT}.

In this paper, motivated by the phenomenological analysis of
ref.\cite{NONUNIV2},
we point out that the combination of both types of
effects in the minimal \su~ naturally gives the physically
desirable non--universal boundary conditions at the GUT scale.
This is a non-trivial prediction of the minimal model which depends
in a crucial way on the presence of the $D$-term contribution
(but not on its actual value) and on the value of the top quark
Yukawa coupling, $h_t$. Two branches of correct solutions are obtained:
one for $h_t$ very close to its quasi-infrared fixed point
and the other one for lower values of $h_t$. With the present uncertainty
on the top quark mass, both solutions may be of phenomenological interest.

The Higgs potential
\be
V=m^2_1H^{\dagger}_1H_1+m^2_2H^{\dagger}_2H_2-m^2_3\left( H^{\dagger}_1
i\tau_2H_2+h.c.\right)\nonumber\\
+\mbox{quartic terms}
\ee
($m^2_i=m^2_{H_i}+\mu^2, m^2_3=B\mu$ where $\mu$ is the supersymmetric
Higgs mixing parameter and $B$ is the corresponding soft term)
has for large $\tb$ values two characteristic features. It follows from the
minimization conditions that
\be
m^2_2\simeq -\frac{M^2_Z}{2}
\label{eq:m2}
\ee
and
\be
m^2_3\simeq\frac{M^2_A}{\tan\beta}\simeq 0 \, ,
\label{eq:m3}
\ee
with
\be
M^2_A\simeq m^2_1+m^2_2>0.
\label{eq:ma}
\ee
Equations
(\ref{eq:m2}) and (\ref{eq:m3})
are the two main constraints on the parameters of the scalar
potential, which are characteristic for large $\tb$ solutions. Combining
(\ref{eq:m2}) and (\ref{eq:ma})
we get
\be
m_1^2-m_2^2 > M_Z^2~.
\label{eq:m1m2mz}
\ee

Let us first discuss the dependence of the low-energy parameters on the
GUT-scale boundary values.  This discussion clearly shows the
need for a special type of non-universality of scalar masses at the GUT
scale.  After that, we demonstrate such a special type of
non-universality can be naturally obtained in SO(10) GUT even with a
universal boundary condition at the Planck scale.

The parameters of the low energy potential are given in terms of their
boundary values at large scale and the RG running. Here we consider
the minimal SUSY-SO(10) model with  matter fields in
16 dimensional representations and the
two Higgs doublets in one representation of dimension 10.
At the GUT scale, after  RG running from the reduced Planck
scale $M_{Pl} \simeq 2.4\cdot 10^{18}$ GeV to the GUT scale
with the SO(10) RG equations, all scalar masses depend on three
parameters $\ovm_{16}, \ovm_{10}$ and $D$ ($D$-term):
\beq
\ovm^2_{H_1,H_2}&=&\ovm^2_{10}+
\left\{
\begin{array}{r}
2\\
-2
\end{array}
\right\}D \, ,
\label{eq:mH0}
\\
\ovm^2_{Q,U,D}&=&\ovm^2_{16}+
\left\{
\begin{array}{r}
1\\
1\\
-3
\end{array}
\right\}D \, .
\label{eq:mF0}
\eeq
The values of the masses at the electroweak scale are obtained
by solving the RG equations of the minimal supersymmetric standard
model with the initial conditons at the GUT scale given by
eqs.(\ref{eq:mH0}) and (\ref{eq:mF0}).
For $Y_t=Y_b=Y\equiv h^2/4\pi$ the approximate solutions
to the 1--loop RG equations read
\cite{CAOLPOWA}:
\beq
m^2_{H_1,H_2}
&\!\!=&\!\!
\left(1-\frac{3}{7}y_Z\right)\ovm^2_{10}
-\frac{6}{7}y_Z \ovm^2_{16}
- c_M \ovM_{1/2}^2 +
\left\{
\begin{array}{r}
2\\
-2
\end{array}
\right\}D + \ldots \, \, ,
\label{eq:mHZ}
\\
m^2_{Q,U,D}
&\!\!=&\!\!
\left(1-\frac{4}{7}y_Z\right)\ovm^2_{16}
-\frac{2}{7}y_Z \ovm^2_{10}
+ c^{Q,U,D}_M \ovM_{1/2}^2
+\left\{
\begin{array}{r}
1\\ 1\\ -3
\end{array}
\right\}D
+\ldots \, \, .
\label{eq:mFZ}
\eeq
$\ovM_{1/2}$ is the gaugino mass at $M_{GUT}$, $c_M \sim \cO (2)$,
$c_M^{Q,U,D} \sim \cO (4)$ and for convenience of analitical solutions
we have introduced the parameter
\be
y_Z=\frac{Y\left(M_Z\right)}{Y_{fZ}}~
\label{eq:yZ}
\ee
where $Y\left(M_Z\right)$ in the well known solution to the MSSM
renormalization group equations
for the Yukawa coupling in the limit of large $\tb$
\cite{IB}, \cite{CAOLPOWA}:
\be
Y\left(M_Z\right)=
\frac
{E_{MSSM}\left(M_Z\right)Y\left(M_{GUT}\right)}
{1+7F_{MSSM}\left(M_Z\right)Y\left(M_{GUT}\right)} \, .
\ee
Here,
$Y_{fZ}$ is the  auxiliary parameter given by the
value of $Y\left(M_Z\right)$ corresponding to the
Landau pole in $Y$ at $M_{GUT}$
\be
Y_{fZ}
=\lim_{Y\left(M_{GUT}\right)\rightarrow\infty}Y\left(M_Z\right)
=\frac{E_{MSSM}\left(M_Z\right)}{7F_{MSSM}\left(M_Z\right)} \, ,
\ee
and $E_{MSSM}$ and $F_{MSSM}$ are functions of the gauge couplings.
The dots in
eqs.(\ref{eq:mHZ}) and (\ref{eq:mFZ})
stand for terms which depend on soft parameter
${\overline A_0}$ at the GUT scale. In this approximation the condition
(\ref{eq:m1m2mz})
gives
\be
m^2_1-m^2_2= 4 D > M^2_Z ~.
\label{eq:m1m1D}
\ee
However, here we have neglected small differences in the running of the two
Higgs masses which follow from the different hypercharges of the
right top and bottom squarks, from the difference in the running
of the bottom and top Yukawa couplings (equal at the GUT scale)
and from the effects due to the $\tau$ lepton Yukawa.
After inclusion of those effects we get
\be
m_1^2 - m_2^2 = a \ovM_{1/2}^2 + c \ovm_0^2 + 4D \, ,
\label{eq:m1m2acD}
\ee
where $\ovm_0$ is an average scalar mass at the GUT scale
(actually $\ovm_0^2 = (\ovm_{10}^2+2 \ovm_{16}^2)/3)$
and the numerical values of the coefficients are
$a \sim |c| \sim \cO (0.1)$ with $c<0$.
The small effects neglected in
eq.(\ref{eq:m1m1D})
but included in
eq.(\ref{eq:m1m2acD})
are resposible for radiative breaking in the case of
universal boundary conditions at the GUT scale
($D=0\,,\,\,\, \ovm_{16}=\ovm_{10}=\ovm_0$)
\cite{CAOLPOWA}.
Then the large $\tb$ solutions must be driven by
large values of $\ovM_{1/2}$:
\be
\ovM_{1/2}>\frac{M_Z}{\sqrt{a}}~,
\qquad
\ovM_{1/2}>\sqrt\frac{|c|}{a}~\ovm_0~,
\label{eq:bigM12}
\ee
and as discussed in
ref.\cite{BOP},
this scenario is strongly disfavoured for several reasons.

It is clear from
eq.(\ref{eq:m1m2acD})
that in the framework of \su~, with $D=d\ovm^2_0$,
qualitatively new solutions become
possible if $c + 4d > 0$, with $\ovM_{1/2}\simeq0$ and
\be
\ovm_0>\frac{M_Z}{\sqrt{c+4d}} \, .
\label{eq:bigm0}
\ee
Thus, with positive $D$,
contrary to the universal case, radiative electroweak
breaking can be driven by soft scalar masses and this pattern
does not depend on  the actual value of the $D$ term as well as on the values
of $\ovm_{16}$ and $\ovm_{10}$. However, as we
shall see, there are important properties of the solutions which
do depend on those masses.

Further properties of the solutions and certain phenomenological
classification of non--universal boundary conditions follows
from the equation (\ref{eq:m2})
\cite{NONUNIV2}.
Since $m^2_2=m^2_{H_2}+\mu^2$, this
equation can be interpreted as an equation for $\mu^2$.
In the universal case large values of $\mu^2$
are needed to cancel large negative values of $m^2_{H_2}$. Now,
with non--universal scalar terms it follows from
eqs.(\ref{eq:mHZ}) and (\ref{eq:m2}) that
\be
\mu^2=
c_M \ovM^2_{1/2}
+c_m\frac{\ovm_{10}^2+\ovm_{16}^2}{2}
+2D
-\frac{M^2_Z}{2}
+\ldots \,\, ,
\label{eq:mu2bar}
\ee
where
\be
c_m=
-\left[
\left( 1-\frac{9}{7}y_Z \right)
+ \left(1 + \frac{3}{7}y_Z \right)
\frac{\ovm_{10}^2 - \ovm_{16}^2}{\ovm_{10}^2 + \ovm_{16}^2}
\right]~ ,
\label{eq:cm}
\ee
and the second term in
eq.(\ref{eq:cm})
is generated by departures from universality. The values of
$\mu^2$ depend, contrary to equation (16), on the pattern of
the deviation from universality in $\ovm_{16}$ and $\ovm_{10}$.
We obtain the following classification
\cite{NONUNIV2}:\\
(A)~$\mu^2>c_M\ovM^2_{1/2}$ for $c_m>0$\\
(B)~$\mu^2<c_M\ovM^2_{1/2}$ for $c_m<0$\\
It is clear that in case (A), generically the values
of $\mu$ remain  large $\mu\gg M_Z$ even in the limit
$\ovM_{1/2}\simeq 0$,
due to the positive correlation with the scalar masses.

In case (B) the parameter $\mu$ can  be arbitrarily small
due to the cancellation betwen the scalar and gaugino contributions in
eq.(\ref{eq:mu2bar})
(with $c+4d>0$ and only then; in the opposite case $\ovM_{1/2}>\ovm_0$, as in
eq.(\ref{eq:bigM12}),
and the cancellation is impossible unless $-c_m>c_M$ which
is very difficult to achieve). Note that for
phenomenological reasons (experimental bounds) we are actually
not interested in the strict limit $\ovM_{1/2}=0$.
Thus, radiative breaking can be driven by
$\ovm_0\gtap \ovM_{1/2}\gtap \mu \simeq \cO (M_Z)$. As shown in
refs.\cite{NONUNIV2}, \cite{BOP},
it is the case (B) which is phenomenologically
acceptable, with higgsino--like lightest chargino and
neutralino. The non--universalities of type (A) suffer from
similar problems as the universal case and are disfavoured by
the combinations of constraints from $\Omega h^2<1$ and $BR(b\rightarrow
s\gamma)$.

The condition $c_m<0$ puts non-trivial contraints on the values of
$\ovm_{16}$ and $\ovm_{10}$ at the GUT scale and on the Yukawa coupling.
In the following we demonstrate
that they are satisfied by the values of the masses obtained by
RG running in the minimal SO(10) model from the Planck scale,
 with universal boundary conditions $M_{1/2}$ and $m_0$ for the
soft gaugino and scalar masses at the Planck scale
(the unbarred quantities denote Planck scale parameters)
and for interesting range of values of the top quark Yukawa coupling.

The set of the relevant \su~ RG equations is as follows:
\beq
&&\frac{d}{dt}\alpha=-b\alpha^2 \, ,
\label{eq:RGEa}
\\
&&\frac{d}{dt}Y=\left(\frac{63}{2}\alpha -14Y\right)Y \, ,
\label{eq:RGEY}
\\
&&\frac{d}{dt}A=\frac{63}{2}\alpha M -14YA \, ,
\label{eq:RGEA}
\\
&&\frac{d}{dt}M=-b\alpha M \, ,
\label{eq:RGEM}
\\
&&\frac{d}{dt}m^2_{16}=-5Y(m^2_{10}+2m^2_{16}+A^2)+\frac{45}{2}\alpha M^2 \, ,
\label{eq:RGEm16}
\\
&&\frac{d}{dt}m^2_{10}=-4Y(m^2_{10}+2m^2_{16}+A^2)+18\alpha M^2  \, .
\label{eq:RGEm10}
\eeq
Here $t=\frac{1}{2\pi}\log\frac{M_{Pl}}{Q}$, $\alpha$ is the SO(10)
gauge coupling, $M$ is the running gaugino mass, $A$ -- the trilinear
soft breaking term and as earlier $Y=h_t^2/4\pi$ is the Yukawa coupling
for the third generation. In eqs.(\ref{eq:RGEa})--(\ref{eq:RGEm10}) we
have explicitly introduced the numerical values for the
$\beta$--functions which depend only on the representation assignment of
the matter and light Higgs fields.  The coefficient $b$ in
eqs.(\ref{eq:RGEa}) and (\ref{eq:RGEM}) depends on the heavy Higgs field
content of the model. We do not have to specify it here as our results
are stable under varying $b$ in the range from 3 to the maximal value
of order 30, corresponding to the Landau pole for the gauge
coupling\footnote{
In models where the heavy Higgs sector correctly
breaks SO(10) to the standard model gauge group without any additional
unwanted massless fields, the beta function is typically $b \geq +7$.
We use $b=+3$ for numerical analyses in this letter as a conservative
choice.  The generated non-universality is larger for larger values of
$b$.  However, as we will describe below, the final result is rather
insensitive on the value of $b$.}.
The solution to those equations read:
\beq
&&\alpha=\frac{\alpha_0}{1+b\alpha_0t}~,
\qquad M=M_{1/2}\frac{\alpha}{\alpha_0}~,
\qquad Y=Y_0\frac{E_G}{1+14Y_0F_G}~,\\
&&E_G=\left(\frac{\alpha_0}{\alpha}\right)^{63/2b}~,~F_G=\int^t_0\!\!E_G
(t')dt'=
\frac{1}{\frac{63}{2}+b}\frac{1}{\alpha_0}
\left[
\left(\aoa\right)^{(63/2b+1)}-1
\right] ~.
\label{eq:EF}
\eeq
The parameters $\alpha_0,~M_{1/2}$ and $Y_0$ are the Planck scale
boundary values of the gauge coupling, gaugino mass and Yukawa
coupling, respectively.

Similarily as for the running from $M_{GUT}$ to $M_Z$
it is convenient to define the parameter
\be
y_G=\frac{Y\left(M_{GUT}\right)}{Y_{fG}}
\label{eq:yG}
\ee
where
\be
Y_{fG}=\lim_{Y_0\rightarrow\infty}Y\left(M_{GUT}\right)
=\frac{E_G\left(M_{GUT}\right)}{14F_G\left(M_{GUT}\right)}~.
\ee
We can then express the parameter $y_Z$ introduced in
eqs.(\ref{eq:yZ}) and (\ref{eq:cm})
in terms of $y_G$:
\be
y_Z=\frac{x y_G}{1 + x y_G}
\label{eq:yZyG}
\ee
where
\be
x=7F_{MSSM}\left(M_Z\right)Y_{fG}~.
\label{eq:x}
\ee
The value of $x$ in eq.(\ref{eq:x}) depends on the scales $M_{GUT}$ and
$M_{Pl}$, on the value of the gauge $\beta$-function coefficient $b$
and on $\alpha_3(M_Z)$ (we take here the attitude that $M_{GUT}$ is determined
by the crossing of $\alpha_1$ and $\alpha_2$ and we allow for a small
mismatch of $\alpha_3$ at that scale
\cite{ChPP}).
For $M_{GUT}=2\cdot 10^{16}
GeV$, $M_{Pl}=2.4\cdot 10^{18} GeV$, $b=3$ and $\alpha_3(M_Z)$ in the range
0.11--0.13, we get $x$ in the range 22--25. It increases (decreases) by 5 for
$b=+30$ (for $M_{Pl}/M_{GUT}$ larger by factor 10).
The condition $c_m<0$,
eq.(\ref{eq:cm}),
now reads:
\be
\frac{\ovm_{10}^2-\ovm_{16}^2}{\ovm_{10}^2+\ovm_{16}^2}
>
\frac{2xy_G-7}{10xy_G+7}~.
\label{eq:m10m16}
\ee
The solutions $c_m=0$ are shown in Fig.1 as the solid curves,
for three different values of $x=15,20,25$. Fig.1 illustrates
the interplay between the GUT scale values of the scalar soft
masses and the Yukawa coupling which is necessary to assure $c_m<0$
for different values of $x$.

As the next step, we solve the
eqs.(\ref{eq:RGEm16}) and (\ref{eq:RGEm10})
and get for the scalar masses:
\beq
\ovm^2_{10}
&\!\!=&\!\!
\left( 1-\frac{12}{14}y_G\right)m^2_0+\frac{4}{14}I~,
\label{eq:m10}
\\
\ovm^2_{16}
&\!\!=&\!\!
\left( 1-\frac{15}{14}y_G\right)m_0^2+\frac{5}{14}I
\label{eq:m16}
\eeq
where
\be
I = a_1A^2_0+a_2A_0M_{1/2}+a_3M^2_{1/2}
\ee
and
\beq
a_1 =
&\!\!-&\!\!
y_G\left(1-y_G\right)
\, ,
\nn\\
a_2 =
&\!\!+&\!\!
y_G \left( 1-y_G \right)
\left[
\left(2 + \frac{63}{b} \right)
\frac
{1-\left(\aoa\right)^{(63/2b)}}
{1-\left(\aoa\right)^{(63/2b+1)}}
-\frac{63}{b}
\right] \, ,
\nn\\
a_3 =
&\!\!-&\!\!
\frac{63}{2b}\left[ \left(\frac{\alpha_0}{\alpha}\right)^2 - 1 + y_G \right]
-y_G \frac{63}{2b}\left(1+\frac{63}{2b}\right)
\frac
{1-\left(\aoa\right)^{(63/2b-1)}}
{1-\left(\aoa\right)^{(63/2b+1)}}
\nn\\
&\!\!-&\!\!
y_G \left( 1-y_G \right)
\left[
\left(\frac{62}{2b}\right)^2
-
\frac{63}{b}\left(1+\frac{63}{2b}\right)
\frac
{1-\left(\aoa\right)^{(63/2b)}}
{1-\left(\aoa\right)^{(63/2b+1)}}
\right]
\nn\\
&\!\!+&\!\!
y_G^2 \left(1+\frac{63}{2b}\right)^2
\left(
\frac
{1-\left(\aoa\right)^{(63/2b)}}
{1-\left(\aoa\right)^{(63/2b+1)}}
\right)^2~.
\eeq

With explicit solutions eqs.(\ref{eq:m10}) and (\ref{eq:m16}) we
can check if the relation (\ref{eq:m10m16}) is indeed fulfilled
in the model. The dashed lines in Fig.1 show the solutions
(\ref{eq:m10}) and (\ref{eq:m16})
for three different values of the ratio $M_{1/2}/m_0$
at the Planck scale and with the values of the other relevant
parameters as specified above eq.(\ref{eq:m10m16}),
with $\alpha_3(M_Z)=0.11$ and $A_0=0$.
It is clear that the solutions to the RG
running from the Planck scale to the GUT scale satisfy the
constraint $c_m<0$ (solid lines) for  values of $y_G$
close to the quasi--IR fixed point or lower than
about 0.2.

This discussion nicely illustrates the role of the boundary values
at $M_{GUT}$ for the scalar masses in obtaining solutions to radiative
breaking in the MSSM with small $\mu$ values. However, since those values
depend on both Planck scale parameters $m_0$ and $M_{1/2}$, it is more
convenient to rewrite
eq.(17) directly in terms of the Planck scale parameters for
effective study of the parameter space. Using
eqs.(32)-(35) we get:
\be
\mu^2=
\left[c_M+\frac{1}{7}\left(3y_Z-2\right) a_3 \right] M_{1/2}^2
+\frac{1}{7}\left[\left(2-3y_G\right)
-9\left(1-y_G\right)\left(1-y_Z\right)\right] m_0^2 + 2D
+\ldots
\label{eq:mu2}
\ee
where the dots stand for $A_0$ dependent terms. The $M_{1/2}$
coefficient remains always positive. The $D$ term must be positive
(see eq.(13) and (14)) and in principle, with $D=dm_0^2$, it should
be included into the $m^2_0$ coefficient. However acceptable solutions
to radiative breaking are obtained already with very small values
of $d$, of order $\cal{O}$(0.01)\footnote{
The maximum possible value of
$d$ can be calculated if one specifies the model completely.  For
instance in the models where the rank is reduced by Higgs fields in 16
and $\overline{16}$ representations, one calculates the difference in
their soft masses $m^2$ and $\bar{m}^2$ using RG equations and
obtains $D = (m^2 - \bar{m}^2)/10$.  $m^2$ and $\bar{m}^2$ can differ
easily by $\cal{O}$(1) because of large group theory factors in SO(10),
and hence $d$ as large as $\cal{O}$(0.1) is possible.}.
At this point it is worth noting
that the negative numerical coefficient $c$ in eq.(14) (which is obtained
from numerical integration of the 1-loop RG equations) goes strictly
to zero for $y_G\rightarrow 1$. This result follows from the structure
of the RG equations and explains why very small positive $d$ is sufficient
to change the pattern of solutions into those of eq.(\ref{eq:bigm0})
(the $A_0^2$ contribution to eq.(\ref{eq:m1m2acD}) is small but
also positive).

Thus, the necessary and sufficient condition for cancellations in
eq.(\ref{eq:mu2})
to be possible is the negative sign of the $m_0^2$ coefficient:
\be
0 >  \left[\left(2-3y_G\right)
-9\left(1-y_G\right)\left(1-y_Z\right)\right] \,.
\label{eq:0}
\ee
By using eq.(\ref{eq:yZyG}) we easily see that this coefficient is
always negative for $x$ in the range (0.6--14.4).
However, for the values of $x$ generic
for the minimal \su~ we obtain non-trivial constraints on the value of the
Yukawa coupling. For $x=22$ the eq.(\ref{eq:0}) is satisfied for $y_G<0.2$
or $y_G>0.6$, in agreement with the results presented in Fig.1.

To study in more detail the $x$ dependence of our result (or equivalently,
for fixed values of $M_{GUT}$ and $M_{Pl}$, its dependence on $\alpha_3(M_Z)$)
and its sensitivity to two-loop corrections in the RG running of the
gauge and Yukawa couplings below $M_{GUT}$, we plot in Fig.2 our two-loop
numerical results as a function of $\alpha_3(M_Z)$. Values of $y_G$ above
and below the band depicted by solid lines satisfy eq.(\ref{eq:0}).
For easy interpretation we also plot the curves of constant top quark
pole masses. One can see that for $\alpha_3(M_Z)=0.11$ \footnote{
Small values of $\alpha_3(M_Z)$ are obtained from the fits to the
electroweak data in the MSSM
\cite{ChP}}
the top quark has to be heavier than 181 GeV
or lighter than 170 GeV, both regions being of phenomenological interest.
For $\alpha_3(M_Z)=0.12$ both bounds move up by about 5 GeV.

Finally, we comment on the prameter space which gives correct radiative
breaking and on the sfermion masses. For instance, for $m_t=182$ GeV,
$\alpha_3(M_Z)=0.11$ and $d=\cO (0.01)$
we get
$\mu^2 = \cO (2) M_{1/2}^2 - \cO (0.05) m_0^2 - \cO (0.01) A_0^2$
and e.g.\ for solutions with $\mu \simeq 100$~GeV and with
$\mu/M_1 \sim 1$ we need $m_0/M_{1/2} = \cO (5)$\footnote{
Contribution from the $A_0^2$ term can lower $m_0$ somewhat.}.
In this simple example the particle spectrum contains a light
pseudoscalar and a higgsino--like chargino, both with masses
below $M_Z$ and within the reach of the Tevatron and LEP2.
It is interesting to note that similar spectrum is predicted from
the best fit to the electroweak data in the framework of the MSSM
\cite{ChP}.

The sfermion masses for the third generation tend to remain
relatively small, too. Combining eqs.(\ref{eq:m10}), (\ref{eq:m16})
with (\ref{eq:mHZ}) and (\ref{eq:mFZ}) we get:
\beq
m_{Q,U,D}^2
\!\!&=&\!\!
\frac{1}{14}
\left[\left( 2-3y_G\right) + 12\left( 1-y_G\right)\left( 1-y_Z\right)
\right] m_0^2
+{\tilde c}_M^{Q,U,D} M_{1/2}^2
+\ldots
\nn\\
\!\!&\simeq&\!\! \cO (0.01) m_0^2 + \cO (4) M_{1/2}^2 + \ldots
\eeq
and analogously for the sleptons,
with ${\tilde c}_M^{L,E} \sim \cO (0.2)$.
In our example, the masses of the third generation squarks are
in the range 200--300 GeV and for sleptons they are $\cO$(100 GeV).
A more detailed study is necessary to check if their contribution
to the breaking of the custodial SU(2) symmetry is consistent
with the electroweak data.

Summary:  In SO(10) SUSY-GUT, a specific pattern of non-universality in
the scalar masses at the GUT-scale is generated by the RGE evolution
from the Planck scale to the GUT scale and the D-term contribution
induced by SO(10) breaking.  This particular pattern of non-universality
can make the $t$-$b$-$\tau$ Yukawa unification phenomenologically viable,
consistent with correct electroweak symmetry breaking,
experimental and cosmological constraints.  The top quark mass is either
below 170~GeV or above 181~GeV if $\alpha_3 (M_Z) = 0.11$.

%%%%%%%%%%%%%%%%%%%%%%%%%%%%%%%%%%%%%%%%%%%%%%%%%%%%%%%%%%%%%%%%%%%%%%
\vskip5mm
\noindent{\bf Acknowledgements}

HM and SP thank the Aspen Center for Physics where the
project was initiated.  HM thanks Lawrence J. Hall for useful
discussions.  HM was supported in part by the
Director, Office of
Energy Research, Office of High Energy and Nuclear Physics, Division of
High Energy Physics of the U.S. Department of Energy under Contract
DE-AC03-76SF00098 and in part by the National Science Foundation under
grant PHY-90-21139.
MO and SP were supported in part by the Polish Committee for Scientific
Research and by the EU grant ``Flavourdynamics''.

%%%%%%%%%%%%%%%%%%%%%%%%%%%%%%%%%%%%%%%%%%%%%%%%%%%%%%%%%%%%%%%%%%%%%

\newpage
\begin{center}
FIGURE CAPTIONS
\end{center}

\begin{itemize}
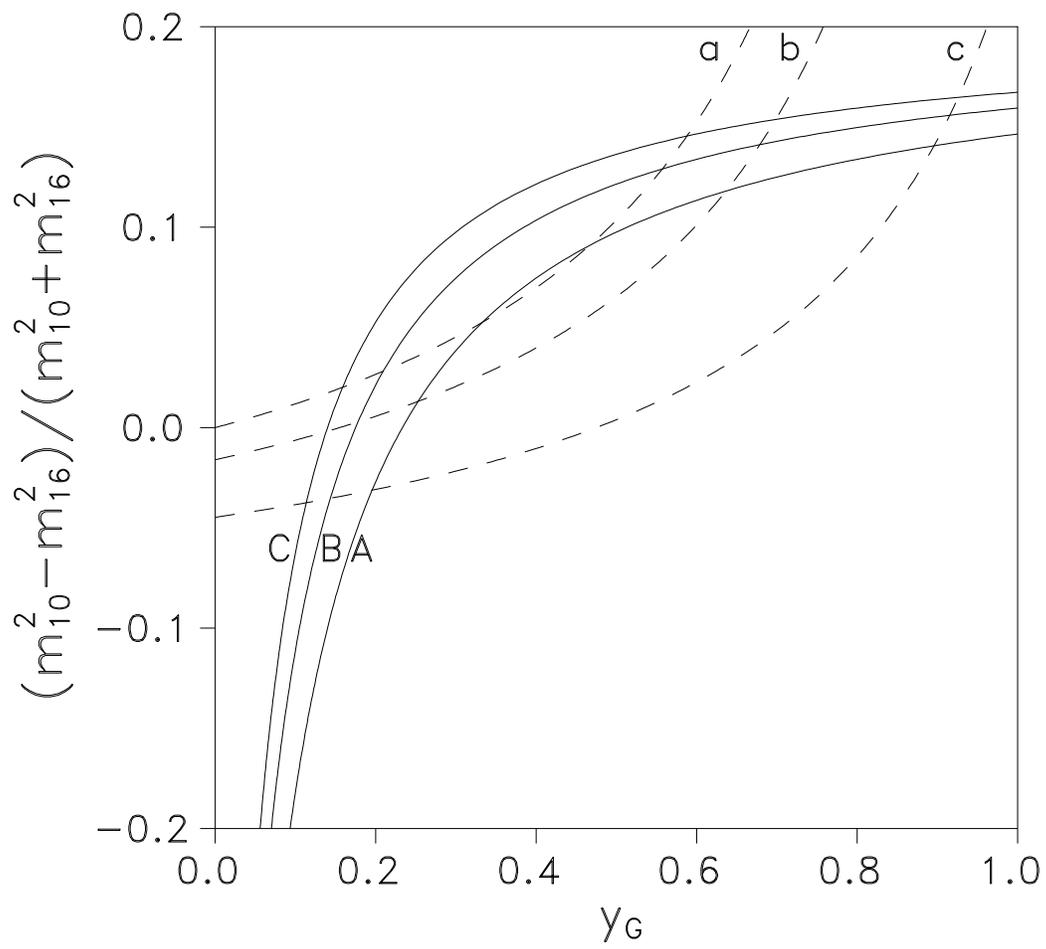
\item[Fig. 1.]
Ratio of masess
$\left(\ovm_{10}^2-\ovm_{16}^2\right)/\left(\ovm_{10}^2+\ovm_{16}^2\right)$
at the GUT scale as a function of the parameter $y_G$ defined in
eq.(\ref{eq:yG}). The solid curves A, B and C are the solutions
to the condition $c_m=0$ for three different values of the parameter
$x=15$, 20 and 25, respectively.
The dashed curves represent results in the SO(10) model for three
fixed values of the ratio $M_{1/2}/m_0=0$, 0.5 and 1.0
(curves a, b and c, respectively) at the GUT scale.
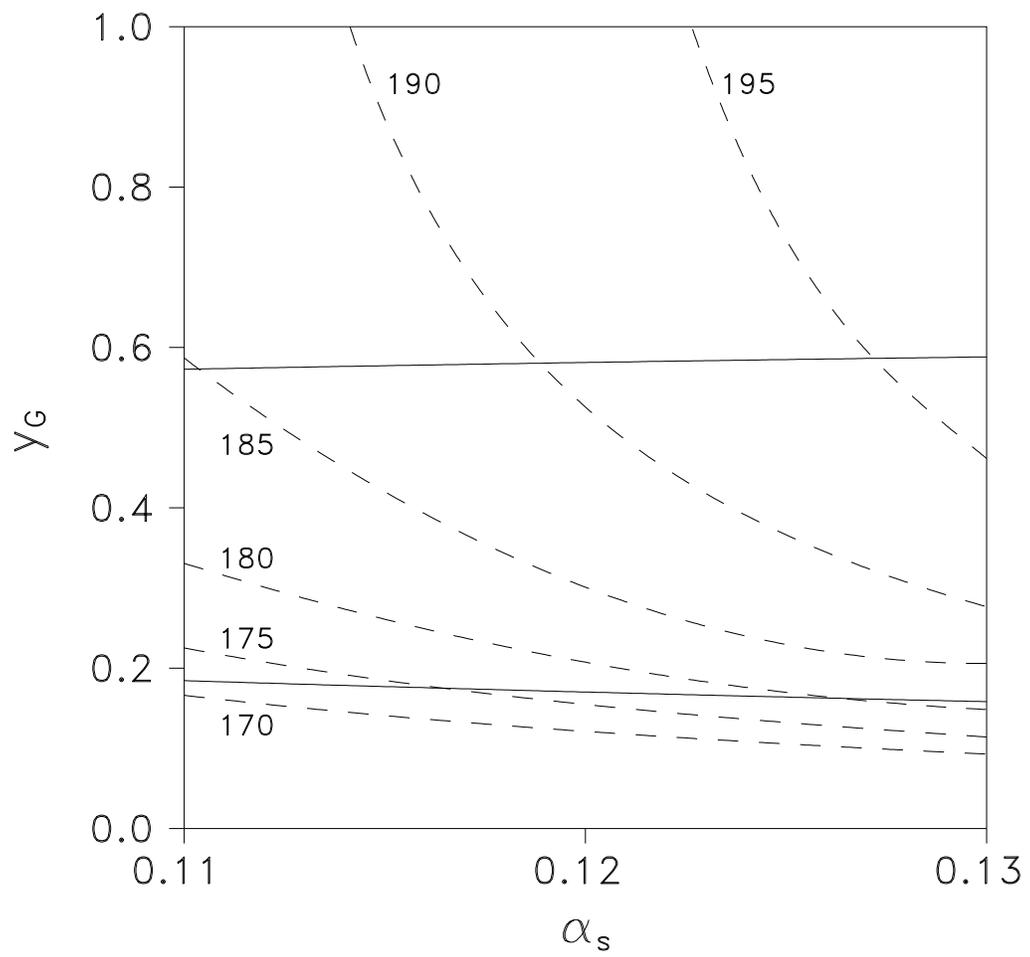
\item[Fig. 2.]
The region in the $y_G$--$\alpha_3(M_Z)$ plane
(outside the band between the solid curves)
in which the condition (\ref{eq:0}) can be satisfied in the minimal
SUSY--SO(10) model. The deshed curves correspond to the fixed values
of the top quark pole mass in GeV. The results were obtained by
integrating numerically the two--loop RG equations below $M_{GUT}$.
\end{itemize}

\end{document}